\documentclass[onecollarge]{svjour3}
\usepackage{graphics,epsfig,graphicx}
\usepackage{bm,xcolor}
\journalname{Few-Body Systems}

\begin{document}

\title{The $ppp$ correlation function with a screened Coulomb potential}

\author{A. Kievsky, E. Garrido, M. Viviani, M. Gattobigio}

\institute{
A. Kievsky \at Istituto Nazionale di Fisica Nucleare, Largo Pontecorvo 3, 56127 Pisa, Italy 
\email{kievsky@pi.infn.it}
\and
E. Garrido \at Instituto de Estructura de la Materia, IEM-CSIC, Serrano 123, E-28006 Madrid, Spain 
\email{e.garrido@csic.sp}
\and 
M. Viviani \at Istituto Nazionale di Fisica Nucleare, Largo Pontecorvo 3, 56127 Pisa, Italy 
\email{viviani@pi.infn.it}
\and
M. Gattobigio \at
 Universit\'e C\^ote d'Azur, CNRS, Institut  de  Physique  de  Nice,
17 rue Julien Laupr\^etre, 06200 Nice, France
}

\maketitle

\begin{abstract}
The correlation function is a useful tool to study the interaction between
hadrons. The theoretical description of this observable requires the
knowledge of the scattering wave function, whose asymptotic part
is distorted when two or more particles are charged. For a system of three 
(or more) particles, with more than two particles asymptotically free and 
at least two of them charged, the asymptotic part of the wave function
is not known in a closed form.
In the present study we introduce a screened Coulomb potential and 
analyze the impact of the screening radius on the correlation function. 
As we will show, when a sufficiently large 
screening radius is used, the correlation function results almost unchanged
if compared to the case in which the unscreened Coulomb potential is used.
This fact allows the use of free asymptotic matching conditions in the solution
of the scattering equation simplifying noticeably the calculation
of the correlation function. As an illustration we discuss the $pp$
and $ppp$ correlation functions.
\end{abstract}

\section{Introduction}

The femtoscopy technique \cite{Wiedemann,Heinz,fem2} is a useful tool to
study the residual strong interaction between hadrons~\cite{lfabb2021}.
In high-energy $pp$ and $p-$nucleus collisions particles are produced and emitted at 
very short relative distances, of the order of the range of the nuclear force. 
Accordingly they interact and the effect of the mutual interaction can be 
captured as a correlation signal in the momentum distributions of the detected particles.
Comparing measurements of correlated particles at low relative energies to theoretical
predictions, it is possible to perform studies of the hadron dynamics. 
Regarding the case in which the system consists in more than two particles, recently, 
the $ppp$ and $pd$ correlation functions have been measured by the 
ALICE Collaboration~\cite{femtoppp,femtopd}. 
In addition, a detailed theoretical
study of the $pd$ correlation function has been done~\cite{pdtheory}.
However, the theoretical description of the 
$ppp$ measurements requires the knowledge of the three-proton scattering wave 
function in the asymptotic region which is not known in a closed form.

Following the recent study of the $ppp$ correlation function where the associate
difficulties regarding the asymptotic configuration has been discussed \cite{theoryppp}, 
here we would like to analyze the
possibility of introducing a screened Coulomb potential in the description of
the correlation function in such a way that free asymptotic boundary conditions
can be used. In general the scattering wave function of two charged particles (we refer to
particles having the same charge) asymptotically never matches the free case. 
No matter the screening radius, when a screened Coulomb potential is
introduced to describe the scattering process of two charged particles
interacting through a short-range potential, the phase-shifts obtained after
matching to the free asymptotic waves are different from those obtained matching to 
the Coulomb wave functions in the unscreened case. In the former case, the phase-shift 
has to be corrected using  
different theoretical contexts, see for example Ref.~\cite{kievsky2010}.
When three particles are asymptotically free and at least two of them are
charged the situation is even more complicated and dedicated techniques
have been developed~\cite{kievsky2001,deltuva2008,yakovlev}.
However, in the case of the correlation function, the source function has a
finite size cutting very fast the scattering wave function. So we expect that
for this case, a sufficient large screening radius will allow to use free
asymptotic boundary conditions without producing appreciable modifications to the
correlation function.

The present paper is organized as follows, we start studying the correlation
function for the $pp$ case using different sizes of the screened Coulomb
potential. After that we make a preliminary study of the $ppp$ correlation 
function using a hypercentral Coulomb force. Then we relax this approximation
and treat the Coulomb potential with and without screening. To this purpose we
made use of the hyperspherical adiabatic basis. The last section
is devoted to the conclusions.

\section{The $pp$ Correlation function with a screened potential}

The $pp$ correlation function results in a convolution of the source function and the
scattering state of two protons. Using the Koonin-Pratt equation~\cite{Koonin,Pratt}
in the center of mass frame it results
\begin{equation}
 C_{pp}(k)=\int d\bm{r} \,  S_{12}(r) |\Psi_s|^2,
 \label{corr}
\end{equation}
where $k$ is the relative momentum of the two protons and $S_{12}(r)$ is the source function, 
depending on the relative distance of the two-protons $r$, introduced
to parameterize the properties of the particle emission process. 
It is defined as a product of two single-particle source functions $S_1({\bf r}_i)$ describing
the spatial distribution of proton $i$ at the position ${\bf r}_i$. The
final form of the source function $S_{12}$ is obtained after integrating on the center of mass 
coordinate of the two protons~\cite{pdtheory,monrow}.
A Gaussian form for the source function $S_{12}$ has been used many times
in the literature, as for example in Ref.~\cite{alicepp} studying the $pp$ and
$p-\Sigma^0$ correlation functions. The same form will be used here as 
discussed below. Moreover, $|\Psi_s|^2$ is the square of the $pp$ 
scattering wave function. If we consider only the Coulomb interaction,
disregarding for the moment the $pp$ nuclear interaction, the scattering wave function for two protons is 
\begin{equation}
	\Psi^0_s= 4\pi\sum_{JJ_z}\sum_{\ell m S_z} i^\ell \frac{F_\ell(\eta,kr)}{kr} (\ell m S S_z|JJ_z)
	 {\cal Y}^{JJ_z}_{\ell S}(\Omega_r) Y^*_{\ell m}(\Omega_k),
	 \label{eq1}
\end{equation}
where $\ell$, $S$, and $J$ are the relative orbital angular momentum, total spin, and total
angular momentum, respectively, with projections $m$, $S_z$, and $J_z$. The angles $\Omega_r$ and $\Omega_k$
indicate the polar and azimuthal angles of the relative coordinate $\bm{r}$ 
and the relative momentum $\bm{k}$.
In the expression above, $F_\ell(\eta,kr)$ is the regular Coulomb function with 
$\eta=e^2 \mu/(\hbar^2k)$ being $\mu$ the reduced mass. The angular-spin functions are
\begin{equation}
	{\cal Y}_{\ell S}^{JJ_z}(\Omega_r)=\sum_{m S_z} (\ell m \, S S_z|JJ_z) Y_{\ell m}(\Omega_r) \chi_{SS_z} \,,
 \label{coup0}
\end{equation}
where $\chi_{SS_z}$ is the spin function coupling the two spin-$\frac{1}{2}$ particles
to $S$=0, 1, and ${Y}_{\ell m}(\Omega_r)$ is a spherical harmonic function. It
should be noticed that due to antisymmetrization requirements the condition
$(-1)^{\ell+S}=1$ is verified, and the spin $S$ is therefore determined by the odd or even character
of the orbital angular momentum $\ell$. Moreover,
for two uncharged particles ($\eta=0$), the Coulomb function, $F_\ell(\eta,kr)$, reduces to the Riccati-Bessel
function $kr j_\ell(kr)$. 

The norm of the scattering wave function, $|\Psi^0_s|^2_\Omega$, is defined as the average over the angular 
coordinates of the square of the wave function, i.e., 
\begin{equation}
    |\Psi^0_s|^2_\Omega = \frac{1}{(4\pi)^2}\int d\Omega_r \int d\Omega_k |\Psi^0_s|^2,
    \label{norm}
\end{equation}
which for the Coulomb case results
\begin{equation}
|\Psi^0_s|^2_\Omega= \frac{2}{N_S} \sum_{[\ell S]} \left(\frac{F_\ell(\eta,kr)}{kr}\right)^2 N_{[\ell S]},
\end{equation}
where $N_S=4$ is the number of spin states and
the factor of 2 has been introduced to impose $|\Psi^0_s|^2_\Omega \rightarrow 1$ as $r \rightarrow\infty$.
The quantity $N_{[\ell S]}$ is the number of allowed states. Without considering a particular symmetry 
it would be
$N_{[\ell S]}=4(2\ell +1)$. However, for antisymmetric states, as it is required
in the case of two protons, we have that
\begin{equation}
N_{[\ell S]}=\left\{ \begin{array}{cc}
             (2\ell +1) & \mbox{if $\ell$ even}  \\
             3 (2\ell +1) & \mbox{if $\ell$ odd} 
                     \end{array} \right.
\end{equation}
leading to the following expression for the norm:
\begin{equation}
|\Psi^0_s|^2_\Omega= \frac{1}{2}\sum_{\ell\equiv{\rm even}} \left(\frac{F_\ell(\eta,kr)}{kr}\right)^2 (2\ell +1) 
	+ \frac{3}{2}\sum_{\ell\equiv{\rm odd}} \left(\frac{F_\ell(\eta,kr)}{kr}\right)^2 (2\ell +1),
	\label{eq:asy0}
\end{equation}
whereas in the no Coulomb case, $\eta$=0, and $F_\ell(\eta, kr)$ has to be replaced
by $krj_\ell(kr)$: 
\begin{equation}
|\Psi^0_s|^2_\Omega= \frac{1}{2}\sum_{\ell\equiv{\rm even}} j_\ell^2(kr)(2\ell +1) 
	+ \frac{3}{2}\sum_{\ell\equiv{\rm odd}} j_\ell^2(kr)(2\ell +1).
	\label{eq:asy1}
\end{equation}

To introduce the short-range interaction between the two protons, Eq.(\ref{eq1}) can be written as:
\begin{equation}
	\Psi_s=4\pi\sum_{JJ_z}\sum_{\ell m S S_z} i^\ell \Psi_{\ell S}^{JJ_z} (\ell m S S_z|JJ_z) Y^*_{\ell m}(\Omega_k),
 \label{eq1b}
\end{equation}
where $\Psi_{\ell S}^{JJ_z}=(kr)^{-1}F_\ell(\eta,kr) {\cal Y}^{JJ_z}_{\ell S}(\Omega_r)$ is the coordinate wave
function of the system with quantum numbers $\ell$, $S$, $J$, and $J_z$.
Considering the nuclear short-range interaction,
the scattering wave function is given by Eq.(\ref{eq1b}) with the coordinate wave function
taking the form
\begin{equation}
    \Psi_{\ell S}^{JJ_z}= \sum_{\lambda} \frac{u^{\lambda}_{\ell}\!(k,r)}{kr} {\cal Y}^{JJ_z}_{\lambda S}(\hat r)\,,
    \label{radeq}
\end{equation}
with $u^{\lambda}_{\ell}\!(k,r)$ the radial solution of the Schr\"odinger
equation (the dependence on $J$ and $S$ is understood). In the general case,
given an incoming channel with orbital angular momentum $\ell$, the short-range interaction can mix it
with an outgoing channel with quantum numbers $\lambda$. Asymptotically
the radial equations in Eq.(\ref{radeq}) is given by
\begin{equation}
    u^\lambda_{\ell}\rightarrow \delta_{\lambda \ell} F_\ell(\eta,kr)+T_{\lambda \ell} {\cal O}_\ell(\eta,kr),
\end{equation}
where 
${\cal O}_\ell(\eta,kr)=G_\ell(\eta,kr)+i F_\ell(\eta,kr)$ describes the outgoing wave function.
The norm of the wave function results in
\begin{equation}
|\Psi_s|^2_\Omega= \frac{1}{2}\sum_{\lambda, \ell\equiv{\rm even}} \left(\frac{u^\lambda_\ell(kr)}{kr}\right)^2 (2\ell +1) 
	+ \frac{3}{2}\sum_{\lambda,\ell\equiv{\rm odd}}
 \left(\frac{u^\lambda_\ell(kr)}{kr}\right)^2  (2\ell +1),
 \label{norm2}
\end{equation}
 which trivially reduces to Eq.(\ref{eq:asy0}) when the short-range interaction is absent.

 To compute the correlation function in Eq.(\ref{corr}) we use a single-particle emission source 
 of a Gaussian form leading to the following two-body source
 function~\cite{pdtheory,monrow}
\begin{equation}
 S_{12}(r)= \frac{1}{8\pi^{3/2}R^3} e^{-(r^2/4R^2)} \ ,
 \label{sour2b}
\end{equation}
 where $R$ is the source radius. In the following a source radius of $R=1.249\,$ fm 
 will be considered.
 Since the source is spherical, the angular integration can be easily performed.
 When the particles interact only through the Coulomb force, the correlation
 function is given by
\begin{equation}
 C^c_{pp}(k)=\frac{1}{4\sqrt{\pi}R^3} \frac{1}{k^2} \int dr e^{-(r^2/4R^2)}
 \left(\sum_{\ell\equiv{\rm even}} F^2_\ell(\eta,kr) (2\ell +1) +
            {3}\sum_{\ell\equiv{\rm odd}} F^2_\ell(\eta,kr) (2\ell +1) \right).
	    \label{eq:cc12}
\end{equation}

To study the effect of screening on the Coulomb potential, we introduce the
following short-range potential
\begin{equation}
	V_{sc}(r)=\frac{e^2}{r}\, e^{-(r/r_{sc})^n} \,,
\end{equation}
where $r_{sc}$ is the screening radius and the parameter $n$ allows for a
sufficient fast cut of the Coulomb potential, the value $n=4$ can be
used~\cite{kievsky2010}. We solve the corresponding Schr\"odinger equation 
for the different partial waves and, since the
above potential does not couple different partial waves, the
resulting radial functions $u_\ell(kr)$ have the following asymptotic form:
\begin{equation}
	u_\ell(kr\rightarrow\infty)\longrightarrow kr[j_\ell(kr)+
	T_{\ell\ell}{\cal O}(kr)]
\end{equation}
with ${\cal O}(kr)=\eta_\ell(kr)+i j_\ell(kr)$. 

The correlation function
calculated using the screened Coulomb potential results
\begin{equation}
C^{sc}_{pp}(k)=\frac{1}{4\sqrt{\pi}R^3} \frac{1}{k^2} \int dr e^{-(r^2/4R^2)}
 \left(\sum_{\ell\equiv{\rm even}} u^2_\ell(kr) (2\ell +1) +
            {3}\sum_{\ell\equiv{\rm odd}} u^2_\ell(kr) (2\ell +1) \right).
	    \label{eq:csc12}
\end{equation}

\begin{figure}[t]
\includegraphics[scale=0.5]{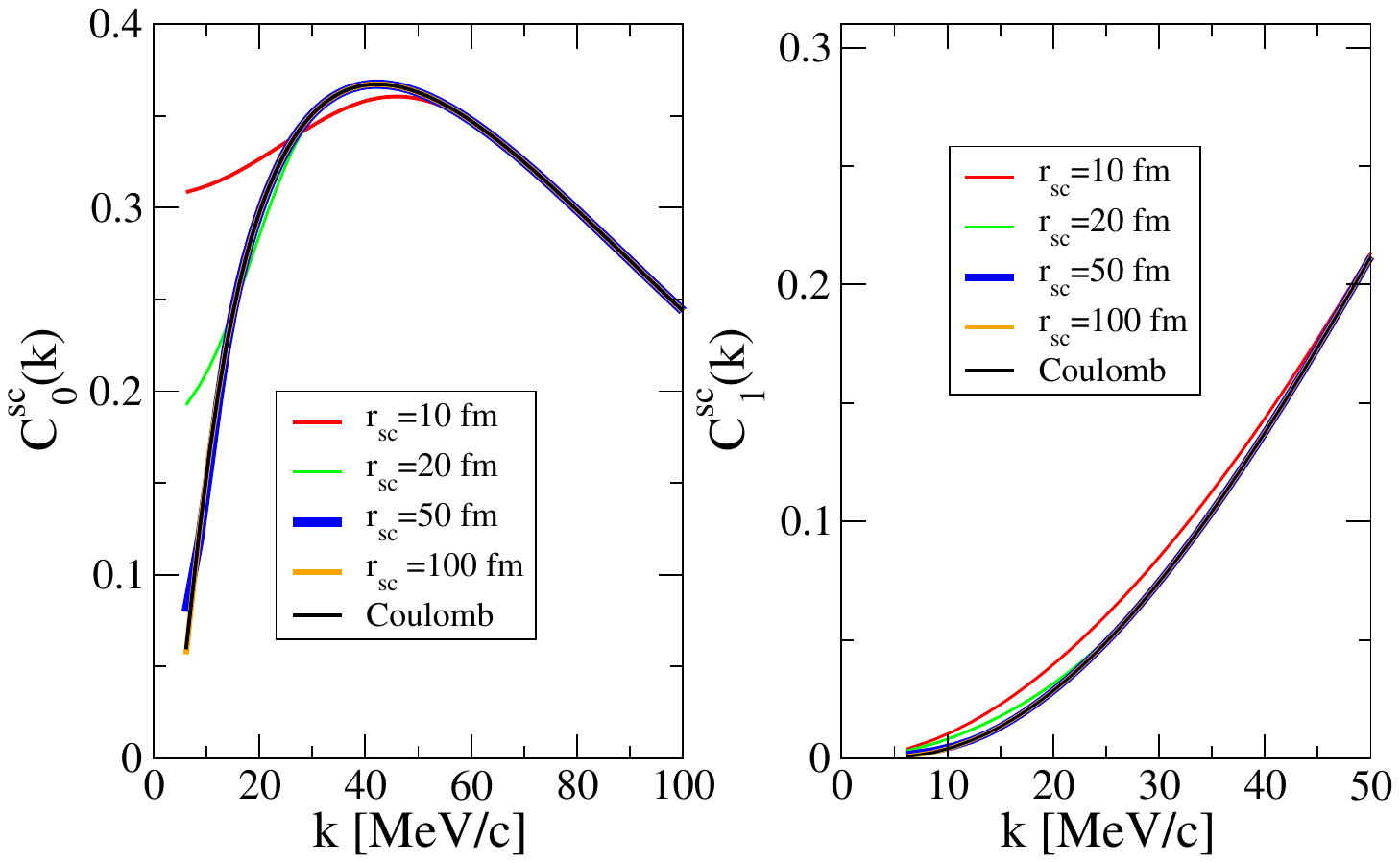}
	\caption{The $\ell=0$ term, $C^{sc}_0(k)$, (left panel) and the
	$\ell=1$ term, $C^{sc}_1(k)$, (right panel) of the screened correlation
	function, $C^{sc}_{pp}(k)$, as a function of the relative momentum $k$,
	calculated for different screening radius $r_{sc}$. The unscreened term is 
	indicated by Coulomb and it is shown as the black curve. }
\label{fig1}
\end{figure}

From Eqs.(\ref{eq:cc12}) and (\ref{eq:csc12}), we can see that both correlation functions, 
$C^c_{pp}=\sum_\ell C^c_\ell$ and $C^{sc}_{pp}=\sum_\ell C^{sc}_\ell$, 
result in a sum of terms, one for each value of $\ell$. 
In Fig.1 we show the first two terms, $C^{sc}_0$ (left) and $C^{sc}_1$ (right) corresponding to
$\ell=0,1$, of the Coulomb screened correlation functions. Screened radii equal to 
$r_{sc}=10,20,50,100\,$ fm have been considered. From the figure,  it is
evident that for values of $r_{sc}> 20\,$ fm the screened correlation function in the
indicated partial waves is almost identical to the Coulomb correlation function. 
Increasing the screening radius the small differences are restricted to lower and lower
relative momentum values. Moreover, for higher partial waves, $\ell>1$, the
terms calculated with or without the screening are almost indistinguishable.

To complete the study of the $pp$ correlation function with screening radius,
we now consider the following $pp$ interaction 
\begin{equation}
	V^{sc}_{pp}(r)=V_0\, e^{-(r/r_0)^2} {\cal P}_0 + V_{sc}(r)\,,
\label{eq:gausspp}
\end{equation}
where the short-range nuclear interaction is modeled by a Gaussian potential
with parameters $V_0=30.45\,$MeV and $r_0=1.815\,$fm, selected to reproduce the
$pp$ scattering length and effective range when used in conjunction with the
unscreened Coulomb potential, $V_C=e^2/r$. In the above formula ${\cal P}_0$ is a projector on spin $S=0$.
The Gaussian $s$-wave potential is a low-energy representation of the
nucleon-nucleon (NN) potential. Its use is justified by the large values
of the NN scattering length which locates the two-nucleon system 
inside the universal window~\cite{kievsky2021}.
A Gaussian representation of the NN interaction has been used many times in recent studies of 
the two-, three- and four-nucleon systems~\cite{higgins2020,higgins2021,tumino2023,gattobigio2019}, 
and even of nuclear matter~\cite{kievsky2018}. It should be noticed that the $pp$
correlation function calculated using the Gaussian potential results extremely
close to that one calculated using a much realistic force as the Argonne
$v_{18}$ interaction.

\begin{figure}[t]
\includegraphics[scale=0.5]{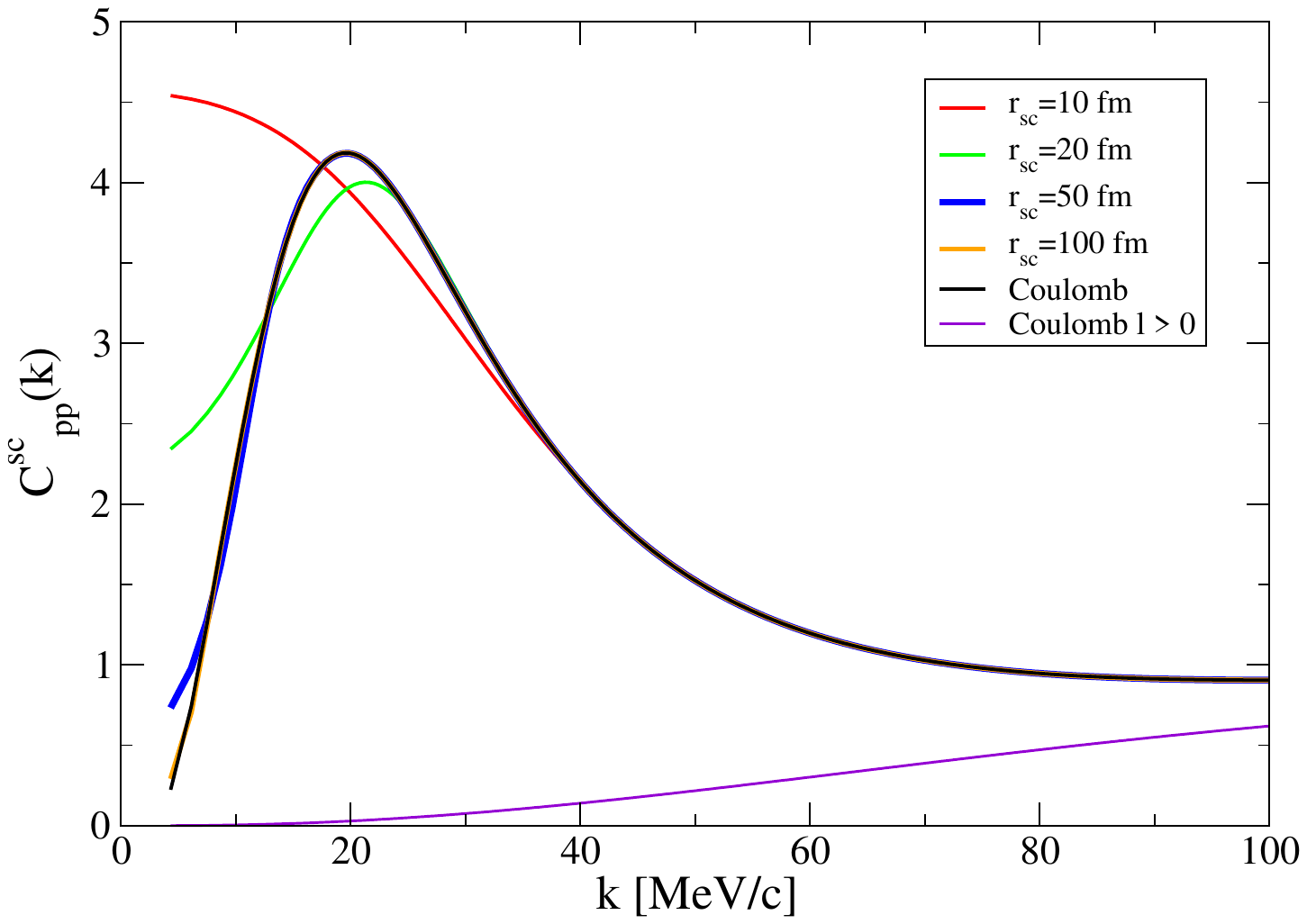}
	\caption{The screened correlation function $C^{sc}_{pp}(k)$ 
        as a function of the relative momentum $k$,
	calculated using the potential of Eq.(\ref{eq:gausspp}) with different screening radius $r_{sc}$. 
	The unscreened term is
        indicated by Coulomb and it is shown as the black curve. The
	contribution of the partial wave with $\ell>0$ are explicitly shown. }
\label{fig2}
\end{figure}

In Fig.2 the screened correlation function is shown for
different screening radii. The radial equation $u_0(r)$ has been computed using
$V^{sc}_{pp}$, as given in Eq.(\ref{eq:gausspp}). The solution
including the Gaussian potential plus the unscreened Coulomb force is shown
in the figure by the black solid line. Finally, the contributions with $\ell>0$
are explicitly shown by the violet curve. From the figure, it is clear that 
a screening radius of $r_{sc}\sim 50\,$fm is large enough to obtain results equivalent
to the case in which the pure Coulomb force is used. This fact, which has little
importance in the case of the correlation function of two particles, could be
very useful when studying correlations of three particles with at least two of
them charged.

\section{The $ppp$ correlation function with a screened Coulomb potential} \label{chap:threebodycase}

The correlation function for three protons can be generalized from the $pp$ case
as
\begin{equation}
C_{ppp}(Q)=\int \rho^5 d\rho \,d\Omega_\rho \,S_{123}(\rho) |\Psi_s|^2,
\label{c123}
\end{equation}
where $S_{123}(\rho)$ is source function, $Q$ is the hypermomentum, $\rho$ is the hyperradius,
$\Omega_\rho$ collects the set of five hyperangles, and $|\Psi_s|^2$ is the
square of the three-proton scattering wave function. 
To describe the $ppp$ wave function we use the Jacobi coordinates 
$\bm{x}=\bm{r}_2-\bm{r}_1$ and $\bm{y}=\sqrt{4/3}\ [\bm{r}_3-(\bm{r}_1+\bm{r}_2)/2]$, 
with $\bm{r}_i$ the position vector of particle $i$. The hyperradius is defined as $\rho=(x^2+y^2)^{1/2}$, and the
five hyperangles consist on the four angles describing the direction of $\bm{x}$ and $\bm{y}$ plus
$\alpha=\arctan(x/y)$. The conjugate momenta are $\bm{k}_x$ and $\bm{k}_y$ and the hypermomentum is
$Q=(k_x^2+k_y^2)^{1/2}$. The five hyperangles in momentum space $\Omega_Q$ are defined 
equivalently to $\Omega_\rho$, but in terms of the momenta $\bm{k}_x$ and $\bm{k}_y$.

Following Ref.\cite{theoryppp} the three-body scattering wave function can be written as 
(see also Refs.\cite{dan04,gar14})
\begin{equation}
    \Psi_s=\frac{(2\pi)^3}{(Q\rho)^{5/2}}  
    \sum_{JJ_z} \sum_{K\gamma}\Psi_{K\gamma}^{J J_z}
    \sum_{M_LM_S} (LM_L SM_S|JJ_z) {\cal Y}_{KL M_L}^{\ell_x\ell_y}(\Omega_Q)^*, 
	\label{3bdcon}
\end{equation}
where the index $\gamma$ refers to the quantum numbers $\{\ell_x,\ell_y,L,s_x,S\}$, 
with $\ell_x$, $\ell_y$ the relative orbital angular momenta associated to 
the coordinates $\bm{x}$ and $\bm{y}$ coupled to total orbital angular 
momentum $L,M_L$. The spin $s_x$ is the spin of the two protons
connected by the $\bm{x}$ coordinate coupled to the spin of the third proton to give
the total spin $S,M_S$. The angular momenta $L$ and $S$ are coupled to the total angular momentum 
of the system $J,J_z$.
The grand-orbital quantum number is $K=2\nu+\ell_x+\ell_y$ (with
$\nu=0,1,2,\cdots$)  and ${\cal Y}_{KLM_L}^{\ell_x \ell_y}$ are hyperspherical
harmonic (HH) functions with well defined angular momentum $L,M_L$. The coordinate
wave functions, $\Psi_{K\gamma}^{J J_z}$, take the form
\begin{equation}
    \Psi_{K\gamma}^{J J_z}=\sum_{K'\gamma'} \Psi^{K'\gamma'}_{K\gamma}(Q,\rho) \Upsilon_{JJ_z}^{K'\gamma'}(\Omega_\rho)\ ,
    \label{3bdcoo}
\end{equation}
where the angular and spin part is
\begin{equation}
    \Upsilon_{JJ_z}^{K\gamma}(\Omega_\rho)=\sum_{M_L M_S} (L M_L S M_S |J J_z) 
    {\cal Y}_{KLM_L}^{\ell_x\ell_y}(\Omega_\rho) \chi_{SM_S}^{s_x}.
    \label{upsilon}
\end{equation}

Similar to the two-body case, we define the norm of the scattering wave function
averaging over the hyperangular coordinates
\begin{equation}
    |\Psi_s|^2_\Omega = \frac{1}{\pi^6} \int d\Omega_\rho \int d\Omega_Q |\Psi_s|^2.
    \label{norm3b}
\end{equation}

The three-proton wave function should be antisymmetric under
particle exchange. To construct the correct symmetry we consider in Eq.(\ref{upsilon}) 
only a selection of HH and spin functions. If the spin of the three protons 
is $S=\frac{1}{2}$, Eq. (\ref{upsilon}) is given by 
\begin{eqnarray}
\Upsilon_{JJ_z}^{K\gamma }(S=\frac{1}{2})= 
\sum_{M_L M_S} (L M_L \frac{1}{2} M_S |J J_z) \sum_\lambda (-1)^\lambda
    \frac{{\cal Y}_{KLM_L}^{\ell_x\ell_y,\bar{\lambda}}(\Omega_\rho) \chi^{\lambda}_{\frac{1}{2}M_S}}{\sqrt{2}}\,,
	\label{eq:mixs}
\end{eqnarray}
where the HH functions ${\cal Y}_{KLM_L}^{\ell_x\ell_y,\bar{\lambda}}$ have 
mixed symmetry of type $\bar{\lambda}$ (see below). They are coupled to the 
three-proton spin state with total spin $S=\frac{1}{2}$ 
\begin{equation}
\chi^{\lambda}_{SS_z}=\sum_{\sigma_x\sigma_y} (\lambda \sigma_x \, \frac{1}{2} \sigma_y |SS_z) \chi_{\lambda \sigma_x} 
\chi_{\frac{1}{2}\sigma_y}\,,
\end{equation}
with $\chi_{\lambda\sigma_x}$ and  $\chi_{\frac{1}{2}\sigma_y}$ the spin
functions of protons 1 and 2, and the one of the third proton, respectively.
The quantum number $\lambda=1,0$ labels the mixed spin
symmetry, symmetric or antisymmetric with respect to
the exchange of particles $1,2$, respectively. With $\bar\lambda$ we indicate
the conjugate symmetry.

When $S=\frac{3}{2}$ the spin part is always symmetric under the exchange of 
protons 1 and 2, then we have
\begin{equation}
    \Upsilon_{JJ_z}^{K\gamma}(S=\frac{3}{2})=\sum_{M_L M_S} (L M_L \frac{3}{2} M_S |J J_z)
    {\cal Y}_{KLM_L}^{\ell_x\ell_y,a}(\Omega_\rho) \chi^1_{\frac{3}{2}M_S}\,,
\end{equation}
where ${\cal Y}_{KLM_L}^{\ell_x\ell_y,a}$, is an antisymmetric HH function
coupled to the symmetric spin $S=\frac{3}{2}$ of the three protons. 
To be noticed that the index $s_x$ is fixed by the symmetry requirements.

When no interaction is considered the hyperradial behavior of the wave function is
\begin{equation}
    \Psi^{K'\gamma'}_{K\gamma}(Q,\rho)=i^K \sqrt{Q\rho}J_{K+2}(Q\rho)\delta_{KK'}\delta{\gamma \gamma'},
    \label{prad}
\end{equation}
with $J_{K+2}(Q\rho)$ a Bessel function of order $K+2$.
Considering antisymmetrization the norm results (see Ref.\cite{theoryppp} for details)
\begin{equation}
|\Psi^0_s|^2_\Omega = \frac{6}{N_S}\frac{2^6}{(Q\rho)^4} \sum_{K} J_{K+2}^2(Q\rho) N_{ST}(K)\ ,
\label{free3n}
\end{equation}
where $N_S=8$ is the number of spin states and the factor $6$ in the numerator
assures that the norm tends to unity as $Q\rightarrow\infty$.
$N_{ST}(K)$ is the number of
antisymmetric states, depending on the grand angular quantum number $K$.
In Ref.\cite{theoryppp} it was shown how to calculate $N_{ST}(K)$ for each value
of $K$. 
For three protons the isospin is $T=3/2$, which is completely symmetric.
The spin function is either of mixed symmetry ($S=\frac{1}{2}$) 
or symmetric ($S=\frac{3}{2}$). 
The mixed symmetry spin states combine with the mixed HH functions to form
antisymmetric states, whereas the symmetric spin state has to be combined with
antisymmetric HH functions. Accordingly the norm results
\begin{equation}
|\Psi^0_s|^2_\Omega = \frac{6}{8}\frac{2^6}{(Q\rho)^4} \sum_{K\ge1} J_{K+2}^2(Q\rho)( N^m_{ST}(K)+4N^a_{ST}(K))\,,
\label{free3na}
\end{equation}
with $N_{ST}^m(K)$ and $N_{ST}^a(K)$ the number of mixed and antisymmetric HH
functions, respectively.  Since the spatially symmetric state is not present the
sum starts with $K=1$.

Here we would like to discuss the case in which the Coulomb force is screened.
As a preliminary step we introduce the hypercentral Coulomb force 
obtained after averaging the bare Coulomb force on the hyperangles
\begin{equation}
V_\mathrm{Coul}(\rho)=\frac{1}{\pi^3}\int d\Omega_\rho \sum_{i<j} \frac{e^2}{r_{ij}} = 
\frac{3(4\pi)^2}{\pi^3}\int d\alpha \sin^2\alpha \cos^2\alpha \frac{e^2}{\rho\cos\alpha}
=\frac{16}{\pi}\frac{e^2}{\rho}.
\label{vcoul}
\end{equation}
This procedure transforms the Coulomb potential interacting among the three
protons into a function depending only on the hyperradius $\rho$. 
The hypercentral Coulomb potential defined above is the first term ($K=0$ term) of the expansion 
of the Coulomb potential in terms of HH
functions (see Ref.\cite{fabre1}). In the case of three protons the Coulomb
potential is symmetric and it has to be expanded in terms of symmetric HH
functions. As the symmetric $K=2$ HH function does not exist the next term of
the expansion is the $K=4$ term. This makes the hypercentral approximation of the
symmetric Coulomb force a good approximation in the calculation of the $ppp$
correlation function (for a more complete treatment of the Coulomb interaction 
see Ref.~\cite{garrido2016}).

When solving the three-proton problem with the hypercentral Coulomb potential
the asymptotic solution is a regular Coulomb function with order $K+\frac{3}{2}$ 
and Sommerfeld parameter $\eta=16 m e^2/(\pi \hbar^2 Q)$, where $m$ is the proton mass.
Therefore, for the $ppp$ system without considering for the moment the nuclear
force between the protons, the norm of the continuum wave function is
formally equal to the case discussed above, Eq.(\ref{free3n}), replacing
$ J_{K+2}(z) \longrightarrow \sqrt{\frac{2}{\pi z}} F_{K+\frac{3}{2}}(z)$.
When this is done, the norm is
\begin{equation}
|\Psi^0_s|^2_\Omega=\frac{96}{\pi} \frac{1}{(Q\rho)^5} \sum_KF_{K+3/2}^2(Q\rho)
( N^m_{ST}(K)+4N^a_{ST}(K)) .
\label{eq:psinorm}
\end{equation}

To compute the correlation function defined in Eq.(\ref{c123}) for the case of
three protons, we consider a source function of the Gaussian type
\begin{equation}
S_{123}(\rho)=\frac{1}{\pi^3\rho_0^6}e^{-(\rho/\rho_0)^2}\ , \label{sour3b}
\end{equation}
with the normalization condition
\begin{equation}
\int S_{123}(\rho) \rho^5 d\rho \,d\Omega_\rho=1.
\end{equation}
The parameter $\rho_0$ of the three-proton source function could be related to $R$,
the parameter of the two-body source function, as $\rho_0=2R$ (see the
discussion of Ref.\cite{theoryppp}). Here, however, we consider the $\rho_0$ 
a parameter of the theory and in the following we use $\rho_0=2\,$fm. 

Without considering any interaction, the free correlation function is defined as
\begin{equation}
C^0_{ppp}(Q)= \frac{6}{8}\frac{2^6}{Q^4\rho_0^6}
\int \rho\, d\rho\, e^{-\frac{\rho^2}{\rho_0^2}}\, \sum_K J^2_{K+2}(Q\rho)\, 
	[N^m_{ST}(K)+4N^a_{ST}(K)].
\label{c0nnn}
\end{equation}
In the case in which the Coulomb force is considered in its hypercentral approximation it results
\begin{equation}
C^{0,c}_{ppp}(Q)= \frac{96}{\pi}\frac{1}{Q^5\rho_0^6}
\int \rho\, d\rho\, e^{-\frac{\rho^2}{\rho_0^2}}\, \sum_K F^2_{K+3/2}(Q\rho) \, 
[N^m_{ST}(K)+4N^a_{ST}(K)]\,.
\label{c0hypc}
\end{equation}

To study the screening we first introduce the screened hypercentral Coulomb
potential
\begin{equation}
	V^{sc}_\mathrm{Coul}(\rho)=\frac{16}{\pi}\frac{e^2}{\rho}
	e^{-(\rho/\rho_{sc})^n}
\label{vhypsc}
\end{equation}
with $\rho_{sc}$ the screening hyperradius and, as in the $pp$ case, the value
$n=4$ will be used. In this way the hypercentral Coulomb potential transforms
into a short-range hypercentral potential which modifies
the hyperradial behavior of the wave function
\begin{equation}
    \Psi^{K'\gamma'}_{K\gamma}(Q,\rho)=u_{K}(Q\rho)\delta_{KK'}\delta{\gamma \gamma'}.
    \label{prad1}
\end{equation}
Since this short-range  hypercentral potential does not couple the different $K$-terms of the
Hamiltonian, the hyperradial functions $u_K$ are therefore solutions of a single equation in the 
hyperradius, and their asymptotic form is then given by:
\begin{equation}
        u_K(Q,\rho\rightarrow\infty) \rightarrow
                i^{K}\sqrt{Q\rho} \left[ J_{K+2}(Q\rho)+ T_{KK} {\cal O}_{K+2}(Q\rho) \right],
 \label{asymus}
\end{equation}
with $T_{KK}$ the $T$-matrix element and 
${\cal O}_{K+2}(Q\rho)=Y_{K+2}(Q\rho)+i J_{K+2}(Q\rho)$ is the outgoing wave function.

With these considerations, the correlation function in the case of the potential
defined in Eq.(\ref{vhypsc}) results
\begin{equation}
	C^{0,sc}_{ppp}(Q)= \frac{6}{8}\frac{2^6}{Q^4\rho_0^6}
\int \rho\, d\rho\, e^{-\frac{\rho^2}{\rho_0^2}}\, \sum_K |u_{K}|^2(Q\rho)  \,
	[N^m_{ST}(K)+4N^a_{ST}(K)].
\label{c0sc}
\end{equation}

\begin{figure}[t]
\includegraphics[scale=0.5]{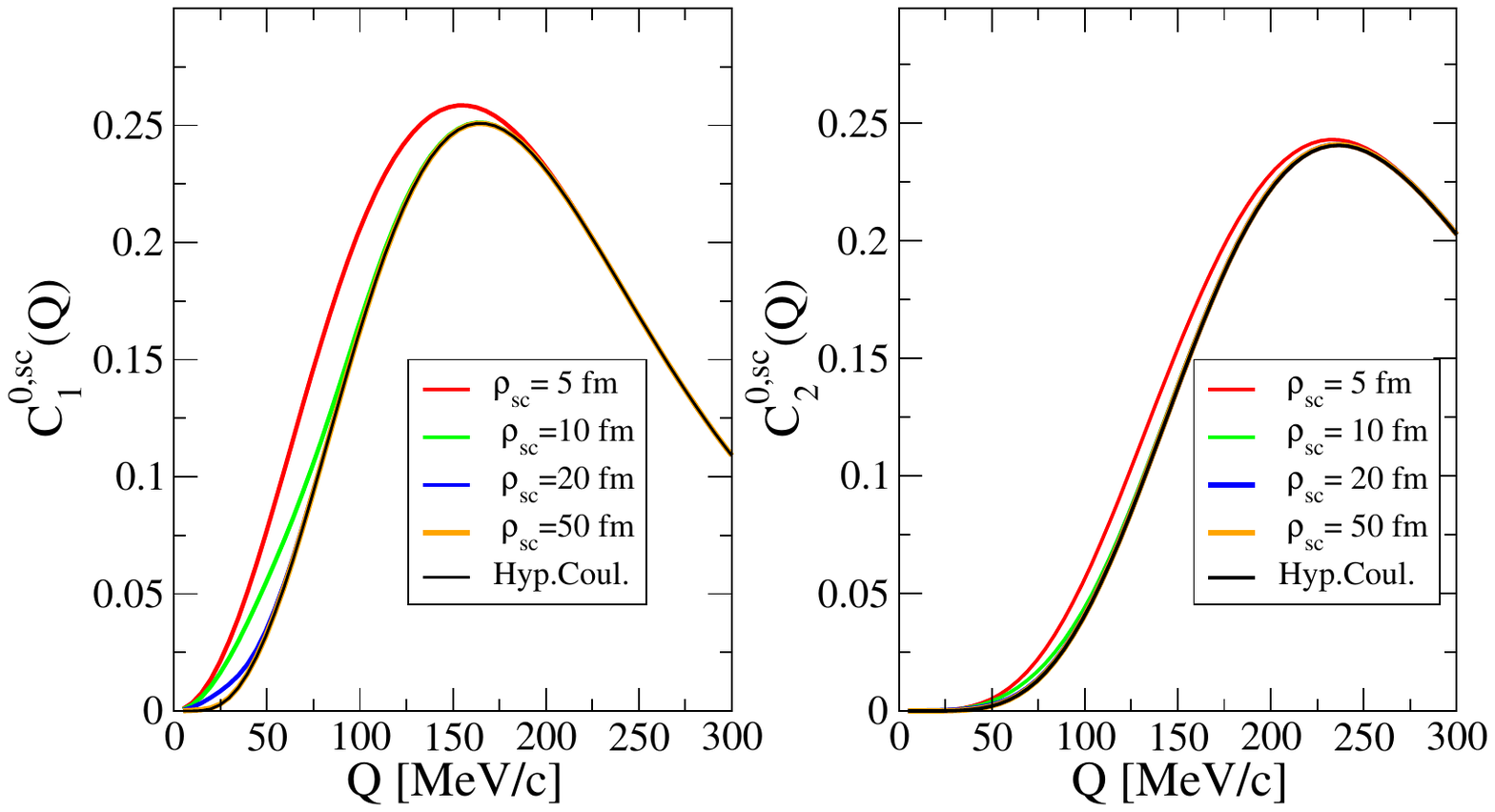}
        \caption{The $K=1$ term, $C^{0,sc}_1(Q)$, (left panel) and the
        $K=2$ term, $C^{0,sc}_2(Q)$, (right panel) of the screened correlation
        function, $C^{0,sc}_{ppp}(Q)$, as a function of the hypermomentum $Q$,
        calculated for different values of the screening hyperradius $\rho_{sc}$. 
	The unscreened term, $C^{0,c}_1(Q)$ and $C^{0,c}_2(Q)$, are 
	indicated by Hyp.Coul. (hypercentral Coulomb) and are shown as the black curves. }
\label{fig3}
\end{figure}

The correlation functions defined in Eqs.(\ref{c0hypc}) and (\ref{c0sc}) result
in a sum of terms labeled by the grand-orbital quantum number $K$,
$C^{0,c}_{ppp}=\sum_K C^{0,c}_K$ and
$C^{0,sc}_{ppp}=\sum_K C^{0,sc}_K$. In Fig.3 we compare the first
two terms of both sums, corresponding to $K=1,2$, using different values of the
screening hyperradius $\rho_{sc}$. We observe that in the case of the first term,
$K=1$, a screening hyperradius of $\rho_{sc}=50\,$fm is already sufficient to
obtain a complete agreement with the unscreened case. In the case of the second
term, $K=2$, a much lower value, around $\rho_{sc}=10\,$fm, is already enough.
Increasing further the values of $K$ the effect of the hypercentral Coulomb
potential vanishes and the contributions to the correlation function with
and without the long-range force are almost equal. Though this analysis could
slightly depend on the size of the source $\rho_0$, this is a consequence of the
short-range character of the source combined with the strong centrifugal barrier
created by the grand-orbital quantum number $K$ which essentially pushes the
effects of the components in the wave function with high values of $K$ to
higher and higher values of $Q$.

To complete the study of the $ppp$ correlation function with a screened Coulomb
potential we calculate the $ppp$ scattering wave function using the potential of
Eq.(\ref{eq:gausspp}) and compare to the case in which the Coulomb force
is not screened. To be precise, we calculate the $ppp$ scattering wave function
using the short-range $pp$ Gaussian potential associated with the Coulomb
and screened Coulomb potential
\begin{eqnarray}
	V_{pp}(r)&=&V_0 \,e^{-(r/r_0)^2} {\cal P}_0 + \frac{e^2}{r}
\label{eq:gaussc} \\
	V^{sc}_{pp}(r)&=&V_0\,e^{-(r/r_0)^2} {\cal P}_0 +
	\frac{e^2}{r}\,e^{-(r/r_{sc})^4} .
\label{eq:gausssc}
\end{eqnarray}

\begin{figure}[h]
\includegraphics[scale=0.5]{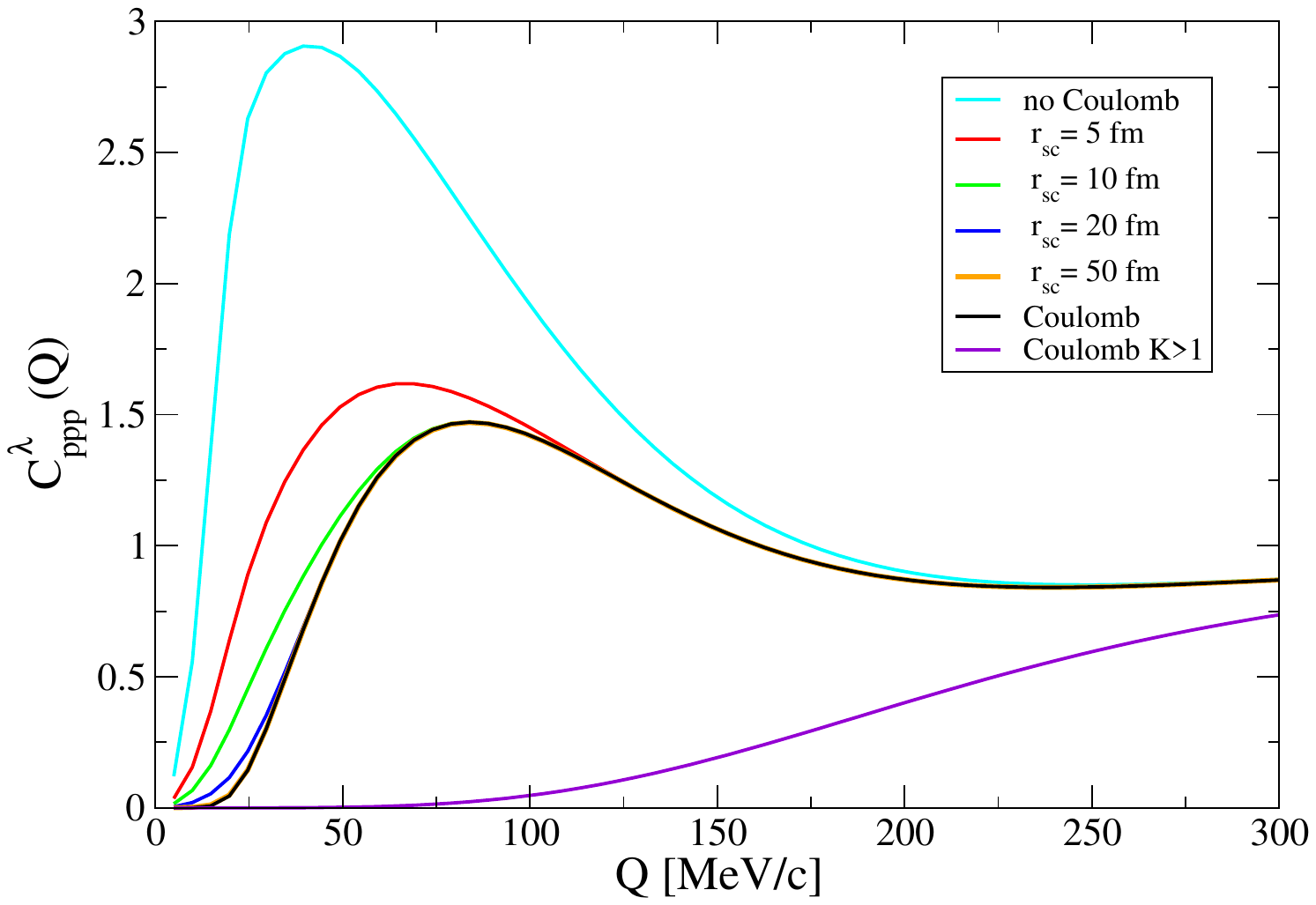}
	\caption{The three-proton correlation function $C^\lambda_{ppp}(Q)$ for
	different screening radii, $r_{sc}$, of the Coulomb interaction. The
	unscreened case is shown as the black curve whereas, for the sake of
	comparison, the no Coulomb case is shown by the cyan curve. The
	contributions of the $K>1$ levels is given by the violet curve.}
\label{fig4}
\end{figure}

To compute the $ppp$ scattering wave function we use the hyperspherical adiabatic (HA)
expansion as described in Refs.\cite{theoryppp,gar14}. Since essentially the
effect of the Coulomb force is more evident in the lowest states we limit the
study of the screening to the first adiabatic channel. All the other channels
will be considered as free. Accordingly the correlation function will be defined
as
\begin{equation}
	C^\lambda_{ppp}(Q)=C^\lambda_1(Q)+\sum_K C^{0,c}_K(Q)\,,
\end{equation}
with $\lambda\equiv c$ we indicate the case in which the adiabatic equations are
solved for the first adiabatic
level using the potential of Eq.(\ref{eq:gaussc}) whereas with $\lambda\equiv sc$ we 
indicate the case in which we solve the adiabatic equations for the first
adiabatic level using the potential of Eq.(\ref{eq:gausssc}) using different
screening radius. In the first and second cases the hyperradial functions will be matched 
to Coulomb and Bessel functions, respectively. In both cases, as indicated in the above
equation, channels with $K>1$ has been calculated using 
Eq.(\ref{c0hypc}). In Fig.\ref{fig4} we show the results where, for the sake of
comparison, we have included the case in which the Coulomb potential is not
considered in the solution of the first adiabatic equation (cyan curve).
Moreover in the figure the contribution  of the $K>1$ channels is indicated by
the violet curve. From the figure we can observe that a screening radius of
$r_{sc}=50\,$ fm is sufficient for an accurate description of the correlation
function.

\section{Conclusions}

In the present work we have studied the effects of the screening of the Coulomb
interaction in the correlation function. Since the correlation function is
mostly measured in the case of charged particles, the correct description of the
asymptotic configuration is an important piece of the theoretical treatment.
However, when more than two charged particles are present in the asymptotic 
configuration, a closed description is not available and different approximations
can be used. The correlation function can be taken as a case study for the
screening. 
We have started studying the $pp$ correlation function and observed that, due to
the finite size of the source, it is possible to screen the Coulomb interaction
without appreciable modifications of the final result, provided that we take
care of the values of the screening radius.
To make contact with
previous studies, in the analysis we have used a source size of $R=1.249\,$ fm,
and we have found that a screening radius $r_{sc}\ge 50\,$fm is large enough for
a good description of the observable, in complete agreement with the case
in which the Coulomb force is taken into account.

The second part of the study regards the $ppp$ correlation function. This case
has some importance due to the considerations explained above. In first place we
have used the Coulomb potential in its hypercentral form. 
We have demonstrated that, similar to the $pp$ case, the correlation function
calculated after screening the hypercentral Coulomb force closely resembles the
case in which screening is not introduced.
Again,
some care is necessary in the choice of the screening hypercentral radius
$\rho_{sc}$, and we have found that a value of $\rho_{sc}\approx
50\,$fm is sufficient large to obtain
a complete agreement with the unscreened calculation. Here the source size was
taken $\rho_0=2\,$fm, a value compatible with previous studies~\cite{theoryppp}.
This preliminary study
is twofold, from one side
the use of the hypercentral force helped to construct
the contributions of channels with values of $K$ sufficiently high to hind the
effects of the nuclear force between the protons. On the other hand it is
useful to perform calculations using free and Coulomb matching
conditions.

In the final part of the study we relax the hypercentral approximation and
consider the Coulomb interaction with and without screening. For this analysis we
use the HA basis limited to the first adiabatic level. In fact, the analysis
done in the $pp$ and $ppp$ cases have shown that the effects of the screening are
more important in the lowest levels. Higher channels are to some extent
protected by the centrifugal barrier and, although some effects of the screening
can be observed as well, in the present study we limit the analysis to the
lowest channel which, in the case of three protons, is associated to the $K=1$
state. To be noticed that using the HA basis the first adiabatic channel includes many
HH states, the $K=1$ state appears asymptotically. The results of the study have
shown that in the computation of the $ppp$ correlation function it is possible
to screen the Coulomb interaction without appreciable modifications of the
observable. This study opens the possibility of studying the three-body correlation function
in the case in which the asymptotic configuration includes more than two
particles and at least two of them are charged.

\begin{acknowledgements}
This work has been partially supported by: Grant PID2022-136992NB-I00 funded by MCIN/AEI/10.13039/501100011033 and
by ERDF A way of making Europe.
\end{acknowledgements}

\end{document}